\documentstyle[12pt]{article}

\begin{document}

\thispagestyle{empty}
 \begin{flushright}
 INP MSU 96-34/441 \\[2mm]
 hep-ph/9610510\\[9mm]
 October 1996
 \end{flushright}
\vspace{30 mm}

\begin{center}
{\Large \bf Some techniques for calculating \\[2mm] 
            two-loop diagrams}
\footnote{
Talk given at the
International Symposium on Radiative Corrections CRAD96
(Cracow, Poland, \\ 1--5 August 1996).
To be published in the proceedings ({\em Acta Physica Polonica}).}
\end{center}

\vspace{10mm}

\begin{center}
Andrei~I.~Davydychev
\end{center}

\vspace{10mm}

\begin{center}
{\em Institute for Nuclear Physics, Moscow State University, \\
119899 Moscow, Russia}
\end{center}

\vspace{15mm}

\begin{abstract}
A brief overview of some recent publications
related to the evaluation of two-loop Feynman diagrams is given.
\end{abstract}
 
\newpage
\setcounter{page}{2}
\setcounter{footnote}{0} 

We shall discuss, very briefly, recent progress in
evaluating certain types of two-loop Feynman diagrams and 
related issues. Some emphasis will be put on the activities
I was involved in, in collaboration with F.A.~Berends,
V.A.~Smirnov, J.B.~Tausk and N.I.~Ussyukina.
Mainly, a collection of ``pointers'' to the papers describing the
relevant methods and algorithms is given. 
The lack of space makes it impossible to include references to all 
papers containing various applications of two-loop calculations,
as well as some problems related to on-shell calculations with
massive particles. 
     
As a rule, the physical case of four dimensions is understood. 
In case of need, the dimensional regularization
\cite{dimreg} is used as a regulator.


\section{Two-loop self-energy diagrams with masses} 


For some special cases, the results were known long time ago.
For example, two-loop diagrams contributing to the photon polarization 
operator in QED were first calculated in 1955 \cite{KS}
(see also in \cite{master1-2}).
Some other special cases were considered in 
refs.~\cite{special}. In all examples, some of the internal
lines were massless and there were just one or two different 
non-zero masses. As a rule, the results were expressible in
terms of trilogarithms $\mbox{Li}_3$, except for one of the QCD 
contributions to the quark selfenergy. 

Indeed, the situation becomes more complicated if one is interested
in a diagram involving a three-particle threshold with all 
massive particles. In this case, there are arguments
\cite{Scharf} that, for a general external momentum $k$, the result
may not be expressible in terms of polylogarithms. 
The simplest example is the so-called ``sunset'' diagram with three 
massive propagators. Recently, it
was considered in a number of papers~\cite{sunset,GvdB}.
The results were expressed either in terms of multiple hypergeometric
series or via one-dimensional integrals.

For more complicated diagrams, e.g.\ for the general two-loop
self-energy diagram with different masses, the representations
in terms of known functions are not available\footnote{In three 
dimensions, an essential progress
has been recently made in ref.~\cite{Rajantie}. 
However, the three-dimensional
case is simpler, due to a very simple (exponential) form of the
massive propagator in the coordinate space.}.
One of the ways is to construct integral representations
and then calculate the result numerically. 
Various approaches to this problem, including some useful
integral representations, were discussed in 
refs.~\cite{Kreimer+BT,Japan,CzKK,Bauberger,GvdB}.
Tensor reduction of two-loop self-energy diagrams
was discussed in \cite{Weigl}. The problem can be
reduced to evaluation of the scalar integrals.

An analytic approach to the calculation of 
two-loop diagrams with different masses consists in constructing the
expansions of these diagrams in different regions. 
For example, when there is no threshold at $k^2=0$, the small momentum
expansion is basically an ordinary Taylor expansion. 
For two-loop diagrams, the general algorithm
for constructing the coefficients of such an expansion was
presented in \cite{DT1}. 
In general, the expansion converges when $k^2$ is below 
the first physical threshold.  
The coefficients of the expansion can be presented in terms of
two-loop vacuum diagrams. Some results for
these diagrams can also be found in \cite{vacuum},
whereas the results for higher powers of propagators
can be obtained using the integration-by-parts technique \cite{ibp}
(this procedure is also described in \cite{DT1}).
The general problem of tensor decomposition of two-loop
vacuum diagrams was discussed in \cite{Tarasov-Pisa,DT3}.
In ref.~\cite{FT}, conformal mapping and Pad\'e approximations were
used to evaluate numerical values beyond the
threshold(s). In ref.~\cite{Tarasov-new}, a modified scheme for 
calculating the coefficients of the expansion was proposed.

The problem of constructing the large momentum expansion of
two-loop self-energy diagrams
(when $k^2$ is larger than all physical thresholds) was considered
in \cite{DST}. 
To do this, the general theory of asymptotic expansions of 
Feynman diagrams \cite{as-ex} (see also in \cite{Smirnov}) was employed.
In four dimensions, the coefficients of the large momentum expansion
of the master two-loop self-energy diagram involve $\ln(k^2)$ and
$\ln^2(k^2)$.  
The lowest term of the expansion of the master
diagram is nothing but the massless integral \cite{6zeta3} 
proportional to $6\zeta(3)$.

Then, the next question is what to do if one is interested in the
threshold behaviour. Dealing with the thresholds may be
unavoidable even in the small momentum expansion, when 
the lowest physical threshold vanishes. In this case, the Taylor
expansion does not work, and one should use a procedure
based on \cite{as-ex}. These algorithms for the zero-threshold
expansion were described in \cite{BDST} (see also in \cite{LvRV}) 
where all possible zero-threshold configurations were considered. 
Moreover, when there is a large mass parameter, one can also use the
large mass expansion \cite{as-ex} to describe the non-zero
threshold behaviour at the small (as compared with the large mass scale)
non-zero thresholds. For two-loop selfenergies, this procedure was 
considered in ref.~\cite{BST}. One should not put any conditions 
on the relative values of $k^2$ and small masses. In this case, 
two-particle-threshold irregularities are in fact described by the 
one-loop two-point diagrams.


\section{Massive and massless two-loop three-point functions}


The problem of calculating the two-loop three-point functions is,
in general, more complicated than the two-point case.
In particular, there are more external invariants
($p_1^2, p_2^2$ and $p_3^2$), and the structure of
singularities is more involved.
Furthermore, there are two basic topologies, the planar one and the
non-planar one, and there is a problem of so-called irreducible
scalar numerators.

On one hand, for three-point functions one can still use approaches 
based on numerical integration of parametric integrals, see e.g.\ 
in \cite{Kreimer3,Japan,CzKK,GvdB}. The approach used in \cite{CzKK}
to calculate the planar vertex function was recently extended to the
non-planar case \cite{FrKK}.     

On the other hand, it is also possible to extend the analytic
approach based on expansion of three-point functions in different
regions. For example, when all three external momenta squared are
small (less than the corresponding thresholds), one can expand
the three-point function 
in a triple Taylor series, using the algorithms similar to ones from
\cite{DT1}. Such an approach was developed in \cite{FT} 
where also conformal mapping and Pad\'e approximations technique
were used. 
An explicit expression for the projectors yielding the coefficients 
of this triple series was given in \cite{DT3}.

Some special cases of three-point functions involving zero thresholds 
were considered in \cite{FST}. As in the two-point case 
\cite{BDST}, the general theory of asymptotic expansions \cite{as-ex}
was applied. Analogous technique can be also used in the case when 
all three external momenta squared are above the corresponding 
thresholds. In particular, one needs the results for the corresponding 
(in general, off-shell) integrals with massless internal lines\footnote{
Note that the results for the (planar and non-planar) diagrams 
with two of the three external momenta being on shell 
(e.g., $p_1^2=p_2^2=0$) were obtained in refs.~\cite{onshell}.}.

The exact result for the off-shell massless planar three-point function
was obtained in ref.~\cite{UD1-2}. In the derivation the ``uniqueness''
technique \cite{uniq} was used. 
This approach was also generalized to the case
of ladder diagrams with an arbitrary number of rungs, see in 
\cite{UD1-2,Bro-ladder}. 
For the two-loop diagram, the result
was presented in terms of polylogarithms up to the fourth order, 
$\mbox{Li}_4$. 
In ref.~\cite{UD3-4} the non-planar diagram was also calculated\footnote{
We note that in refs.~\cite{CzKK,FrKK} the results of \cite{UD1-2,UD3-4}
were used to check the massless limit of the numerical programs.},
as well as the cases when some of the propagators are shrunk.
The result for the non-planar diagram was presented in terms of
the square of an expression involving dilogarithms.
The problem of irreducible numerators (i.e.\ scalar numerators which 
cannot be cancelled against the denominators) was also considered in
\cite{UD3-4} and some results for the integrals with such numerators
were obtained. A complete solution to this problem should also
include an efficient algorithm for calculating the integrals
with higher integer powers of the propagators and irreducible numerators.   
 
\vspace{5mm}

{\bf Acknowledgements.} I am grateful to all organizers of CRAD96,
especially to S.~Jadach, M.~Je\.{z}abek and Z.~W\c{a}s, 
for their hospitality. This work was partly supported by the EU grant
INTAS-93-0744 and by the RFBR grant 96-01-00654.


\begin{thebibliography}{99}

\bibitem{dimreg} G.~'t~Hooft and M.~Veltman,
    {\em Nucl.Phys.} B44 (1972) 189;\\
    C.G.~Bollini and J.J.~Giambiagi, {\em Nuovo Cim.} 12B (1972) 20.

\bibitem{KS} G.~K\"all\'en and A.~Sabry,
{\em Dan.~Mat.~Fys.~Medd.} 29, No.17 (1955) 1.

\bibitem{master1-2} D.J.~Broadhurst, {\em Phys. Lett.} B101 (1981) 423, \\
T.H.~Chang, K.J.F.~Gaemers and W.L.~van Neerven, {\em Nucl. Phys.} 
  B202 (1982) 407; \\
A.~Djouadi, {\em Nuovo Cim.} 100A (1988) 357; \\
B.A.~Kniehl, {\em Nucl. Phys.} B347 (1990) 86. 

\bibitem{special} D.J.~Broadhurst,
{\em Z.~Phys.} C47 (1990) 115; \\
A.V.~Kotikov, {\em Phys.~Lett.} B254 (1991) 158;
{\em Mod. Phys. Lett.} A6 (1991) 677; \\
D.J.~Broadhurst, J.~Fleischer and O.V.~Tarasov,
{\em Z.~Phys.} C60 (1993) 287; \\
R.~Scharf and J.B.~Tausk,
{\em Nucl.~Phys.} B412 (1994) 523; \\
D.T.~Gegelia, K.Sh.~Japaridze and K.Sh.~Turashvili,
{\em Teor. Mat. Fiz.} 101 (1994) 225.

\bibitem{Scharf} R.~Scharf, Diploma Thesis, W\"urzburg, 1991; 
                         Doctoral Thesis, W\"urzburg, 1994.

\bibitem{sunset} F.A.~Berends, M.~B\"{o}hm, M.~Buza and R.~Scharf,
{\em Z.~Phys.} C63 (1994) 227; \\
F.A.~Lunev, {\em Phys. Rev.} D50 (1994) 7735; \\
P.~Post and J.B.~Tausk, {\em Mod. Phys. Lett.} A11 (1996) 2115.

\bibitem{GvdB} A.~Ghinculov and J.J. van der Bij, {\em Nucl. Phys.}
             B436 (1995) 30.

\bibitem{Rajantie} 
A.K.~Rajantie, Helsinki preprint HU-TFT-96-22
            (hep-ph/9606216).

\bibitem{Kreimer+BT} D.~Kreimer, {\em Phys.~Lett.} B273 (1991) 277; \\
F.A.~Berends and J.B.~Tausk, {\em Nucl.~Phys.} B421 (1994) 456.

\bibitem{Japan} J.~Fujimoto, Y.~Shimizu, K.~Kato and Y.~Oyanagi, 
   KEK preprint 92-213.

\bibitem{CzKK} A.~Czarnecki, U.~Kilian and D.~Kreimer,
{\em Nucl.~Phys.} B433 (1995) 259.

\bibitem{Bauberger} S.~Bauberger, F.A.~Berends, M.~B\"{o}hm and M.~Buza,
{\em Nucl.~Phys.} B434 (1995) 383;\\
S.~Bauberger and M.~B\"{o}hm, {\em Nucl.~Phys.} B445 (1995) 25.

\bibitem{Weigl} G.~Weiglein, R.~Scharf and M.~B\"{o}hm,
{\em Nucl.Phys.} B416 (1994) 606; \\
D.~Kreimer, {\em Mod. Phys. Lett.} A9 (1994) 1105.

\bibitem{DT1} A.I.~Davydychev and J.B.~Tausk,
{\em Nucl.~Phys.} B397 (1993) 123.

\bibitem{vacuum} 
J.J.~van der Bij and M.~Veltman, {\em Nucl.~Phys.} B231 (1984) 205; \\
F.~Hoogeveen, {\em Nucl.~Phys.} B259 (1985) 19; \\
J.J. van der Bij and F.~Hoogeveen, {\em Nucl.~Phys.} B283 (1987) 477; \\
C.~Ford, I.~Jack and D.R.T.~Jones, 
{\em Nucl.Phys.} B387 (1992) 373; \\
A.I.~Davydychev and J.B.~Tausk, {\em Phys. Rev.} D53 (1996) 7381.

\bibitem{ibp} F.V.~Tkachov, {\em Phys. Lett.} B100 (1981) 65; \\
     K.G.~Chetyrkin and F.V.~Tkachov, {\em Nucl. Phys.} B192 (1981) 159.

\bibitem{Tarasov-Pisa} 
O.V.~Tarasov, in: {\em New Computing Techniques in Physics Research IV}
(World Scientific, Singapore, 1995), p.~161 (hep-ph/9505277). 

\bibitem{DT3} A.I.~Davydychev and J.B.~Tausk, 
     {\em Nucl. Phys.} B465 (1996) 507.

\bibitem{FT} J.~Fleischer and O.~V.~Tarasov,
{\em Z.~Phys.} C64 (1994) 413.

\bibitem{Tarasov-new} 
O.V.~Tarasov, Preprints DESY-96-099 (hep-ph/9606238) 
and DESY-96-068 (hep-th/9606018). 

\bibitem{DST} A.I.~Davydychev, V.A.~Smirnov and J.B.~Tausk,
{\em Nucl.~Phys.} B410 (1993) 325.

\bibitem{as-ex}
S.G.~Gorishny, S.A.~Larin and F.V.~Tkachov, 
   {\em Phys.~Lett.} B124 (1983) 217; \\
G.B.~Pivovarov and F.V.~Tkachov, Preprints INR P-0370, $\Pi$-459
   (Moscow, 1984); 
   {\em Int.~J.~Mod.~Phys.} A8 (1993) 2241;\\
S.G.~Gorishny and S.A.~Larin, {\em Nucl. Phys.} B283 (1987) 452; \\
K.G.~Chetyrkin, {\em Teor. Mat. Fiz.} 75 (1988) 26; 76 (1988) 207; \\
                      Preprint MPI-PAE/PTh 13/91 (Munich, 1991);\\
S.G.~Gorishny, {\em Nucl. Phys.} B~319 (1989) 633;\\ 
V.A.~Smirnov, {\em Commun. Math. Phys.} 134 (1990) 109.

\bibitem{Smirnov} V.A.~Smirnov,
{\em Renormalization and asymptotic expansions} 
(Birk\-h\"{a}user, Basel, 1991); 
{\em Mod. Phys. Lett.} A10 (1995) 1485.

\bibitem{6zeta3} J.L.~Rosner, {\em Ann. Phys.} 44 (1967) 11; \\
K.G.~Chetyrkin, A.L.~Kataev and F.V.~Tkachov, 
{\em Nucl.~Phys.} B174 (1980) 345.

\bibitem{BDST} F.A.~Berends, A.I.~Davydychev, V.A.~Smirnov
 and J.B.~Tausk, {\em Nucl.~Phys.} B439 (1995) 536.

\bibitem{LvRV} S.A.~Larin, T. van Ritbergen and J.A.M.~Vermaseren,
{\em Nucl.~Phys.} B438 (1995) 278.

\bibitem{BST} F.A.~Berends, A.I.~Davydychev and V.A.~Smirnov, 
 {\em Nucl.Phys.} B478 (1996) 59; 
 Leiden preprint INLO-PUB-5/96 (hep-ph/9606244), 
 to appear in {\em Nucl.~Phys.} B (Proc.~Suppl.).

\bibitem{Kreimer3} D.~Kreimer, {\em Phys.~Lett.} B292 (1992) 341.

\bibitem{FrKK} A.~Frink, U.~Kilian and D.~Kreimer,
Mainz preprint MZ-TH/96-30 (hep-ph/9610285).

\bibitem{FST} J.~Fleischer, V.A.~Smirnov and O.V.~Tarasov,
Bielefeld preprint BI-TP-95-39 (hep-ph/9605392).

\bibitem{onshell} R.J.~Gonsalves, {\em Phys. Rev.} D28 (1983) 1542; \\
    W.L. van Neerven, {\em Nucl. Phys.} B268 (1986) 453; \\
    G.~Kramer and B.~Lampe, {\em J. Math. Phys.} 28 (1987) 945.

\bibitem{UD1-2} N.I.~Ussyukina and A.I.~Davydychev, {\em Phys. Lett.}
    B298 (1993) 363; B305 (1993) 136.

\bibitem{uniq} A.N.~Vassiliev, Yu.M.~Pis'mak and Yu.R. Honkonen,
   {\em Teor. Mat. Fiz.} 47 (1981) 291; \\
N.I.~Ussyukina, {\em Teor. Mat. Fiz.} 54 (1983) 124; \\
D.I.~Kazakov, {\em Phys. Lett.} B133 (1983) 406. 

\bibitem{Bro-ladder} D.J.~Broadhurst, {\em Phys. Lett.} B307 (1993) 132.
  
\bibitem{UD3-4} N.I.~Ussyukina and A.I.~Davydychev, {\em Phys. Lett.}
   B332 (1994) 159; B348 (1995) 503.

\end{thebibliography}
\end{document}